\documentclass[lnicst]{svmultln}
\usepackage{graphicx}
\usepackage{caption}
\usepackage{makeidx}  % allows for indexgeneration
\begin{document}
\mainmatter              % start of the contribution
\title{\textbf{A Framework for Accurate Drought Forecasting System Using Semantics-Based Data Integration Middleware}}
\titlerunning{Semantics-Based Data Integration Middleware} 
\author{Adeyinka K. Akanbi \and Muthoni Masinde}

\authorrunning{A.K. Akanbi \& Muthoni Masinde}   % abbreviated author list (for running head)
\tocauthor{Adeyinka K. Akanbi, Muthoni Masinde}
\institute{Department of Information Technology\\Central University of Technology, Free State, South Africa\\
\email{{aakanbi,emasinde}@cut.ac.za}%\\ WWW home page:
%\texttt{http://users/\homedir iekeland/web/welcome.html}
%\and
%Universit\'{e} de Paris-Sud,
%Laboratoire d'Analyse Num\'{e}rique, B\^{a}timent 425,\\
%F-91405 Orsay Cedex, France}
}
\maketitle              % typeset the title of the contribution
% \index{Ekeland, Ivar} % entries for the author index
% \index{Temam, Roger}  % of the whole volume
% \index{Dean, Jeffrey}

\begin{abstract}        % give a summary of your paper
Technological advancement in Wireless Sensor Networks (WSN) has made it become an invaluable component of a reliable environmental monitoring system; they form the 'digital skin' through which to 'sense' and collect the context of the surroundings and provides information on the process leading to complex events such as drought. However, these environmental properties are measured by various heterogeneous sensors of different modalities in distributed locations making up the WSN, using different abstruse terms and vocabulary in most cases to denote the same observed property, causing data heterogeneity. Adding semantics and understanding the relationships that exist between the observed properties, and augmenting it with local indigenous knowledge is necessary for an accurate drought forecasting system. In this paper, we propose the framework for the semantic representation of sensor data and integration with indigenous knowledge on drought using a middleware for an efficient drought forecasting system.

\end{abstract}
\section{Introduction}

The application of Semantic Technology for drought forecasting is a growing research area. Our work investigates the semantic representation and integration of measured environmental entities with the local Indigenous knowledge (IK) using an ontology to allow reasoning and generate inference based on their interrelationship.  We present a proposed model which outline our research directions, \cite{akanbi2015towards} provides a further overview of the framework towards an accurate drought forecasting system.

In terms of negative impacts, droughts are currently ranked \footnote{The ranking is based on severity, length of event, total area affected, total loss of life, total economic loss, social effect, long-term impacts, suddenness and frequency\cite{chester1993natural}.}  number one (CRED 2012). Compared to other natural disasters such as floods, hurricanes, earthquakes and epidemics, droughts are very difficult to predict; they creep slowly and last longest. The complex nature of droughts onset-termination has made it acquire the title "the creeping disaster" \cite{mishra2010review}. The greatest challenge is designing a framework which can track information about the 'what', 'where' and 'when' of environmental phenomena and the representation of the various dynamic aspects of the phenomena \cite{peuquet1995event}. The representation of such phenomena requires better understanding of the 'process' that leads to the 'event'. For example, a \textit{soil moisture sensor} provides sets of values for the observed property \textit{soil moisture}. The measured property can also be influenced by the \textit{temperature heat index} measured over the observed period. This makes accurate prediction based on these sensor values almost impossible without understanding the semantics and relationships that exist between this various properties. Hypothetically, drought prediction tools could be used to establish precise drought development patterns as early as possible and provide sufficient information to decision-makers to prepare for the droughts long before they happen. This way, the prediction can be used to mitigate effects of droughts.

The technological advancement in Wireless Sensor Networks (WSN) has facilitated its use in monitoring environmental properties irrespective of the geographical location. In their (WSNs) current implementation, these properties are measured using heterogeneous sensors that are mostly distributed in different locations. Further, different abstruse terms and vocabulary in most cases are used to denote the same observed property, thereby leading to data heterogeneity. Moreover, research \cite{mugabe2010use}, \cite{masinde2011itiki} on indigenous knowledge (IK) on droughts has pointed to the fact that IK on living and non-living things e.g., \textit{sifennefene worms}, \textit{peulwane birds}, \textit{lehota frogs} and plants like \textit{mutiga tree}, \textit{mothokolo tree} etc can indicate drier or wetter conditions, which can imply likely occurrence of drought event over time \cite{sillitoe1998development}. This scenario shows that environmental events can be inferred from sensors data augmented with IK, if proper semantic is attached to it based on some set of indicators. Therefore, a semantics-based data integration middleware is required to bridge the gap between heterogeneous sensor data and IK for an accurate drought forecasting and prediction system.

\section{Problem Statements}
The following problems were identified as a major bottleneck for the utilization of semantic technologies for drought forecasting:
\\ 
\textit{The current lack of ontology based middleware for the semantic representation of environmental process:} Ontological modeling of key concepts of environmental phenomena such as object, state, process and event, ensures the drawing of accurate inference from the sequence of processes that lead to an event. Presently, what is currently missing is an environmental ontology with well-defined vocabularies that allow explicit representation of the process, events and also attach semantics to the participants in the environmental domain.
\\
\textit{Lack of semantic integration of heterogeneous data sources with indigenous knowledge for an accurate environmental forecasting:}  Studies reveal that over 80\% of farmers in some parts of Kenya, Zambia, Zimbabwe and South Africa rely on Indigenous knowledge forecasts (IKF) for their agricultural practices \cite{masinde2011itiki}. An IoT-based environmental monitoring system made up of interconnected heterogeneous weather information sources such as sensors, mobile phones, conventional weather stations, and indigenous knowledge could improve the accuracy of environmental forecasting.
\\
\textit{Lack of IoT-based drought forecasts communication and dissemination channels:} There is a lack of effective dissemination channels for drought forecasting information. For example, the absence of smart billboards placed at strategic location and smart phones. The output channels would ensure farmers have access to drought forecasting information know the spatial distribution of a \textit{drought vulnerability index}.

\section{Research Questions}
\textit{To what extent does the adoption of knowledge representation and semantic technology in the development of a middleware enable seamless sharing and exchange of data among heterogeneous IoT entities?}
\\
Several standards have been created to cope with the data heterogeneities. Examples are the Sensor Markup Language (SensorML) \footnote{http://www.opengeospatial.org/standards}, WaterML, and American Federal Geographic Data (FGDC) Standard \footnote{https://www.fgdc.gov/metadata}. However, these standards provide sensor data to a predefined application in a standardized format, and hence do not generally solve data heterogeneity. Semantic technology solves this by representing data in a machine readable language such as Resource Description Framework (RDF) and Ontology Web Language (OWL), for seamless data sharing irrespective of the domain.
\\
\textit{What are the main components of an implementation framework/architecture that employs the middleware to implement an IoT-based Drought Early Warning Systems (DEWS)?}
\\
The existence of ontology with well-defined vocabularies that allows an explicit representation of process and events; the representation and integration of the inputs in machine-readable formats, the availability of a reasoning engine (\textit{CEP Engine}) that generates inference based on input parameters.

\section{Methodology}
The proposed semantic middleware is a software layer composed of a set of various sub-layers interposed between the application layer and the physical layer. It incorporates interface protocols, which liaise with the storage database in the cloud for downloading the semi-processed sensory reading to be represented based on the ontology through a mediator device as shown in figure 3\cite{akanbi2015towards}. An environmental process-based ontology is required to overcome the problems associated with the dynamic nature of environmental data and the data heterogeneities. The study proposes to use DOLCE top-level ontology for the modelling of the foundational entities needed to represent the dynamic phenomena. Information from the sensor data streams is integrated with indigenous knowledge using a Complex Events Processing (\textit{CEP}) engine as proposed in figure 1. This will serve as the reasoning engine for inferring patterns leading to drought, based on a set of rules derived from indigenous knowledge of the local people on drought. Figure 2 depicts the overview of the middleware architecture. The domain of this particular case study is Free State Province, South Africa - an ongoing research project by AfriCRID\footnote{http://africrid.com/}, Department of Information Technology, Central University of Technology, Free State.

\begin{figure}

\minipage{0.52\textwidth}
  \includegraphics[width=\linewidth]{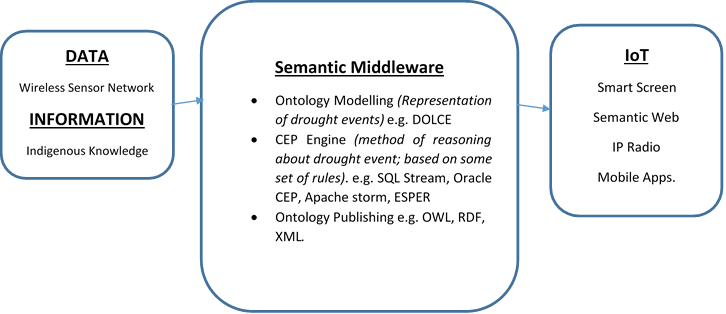}
\caption{The semantic middleware integration framework}
\endminipage\hfill
\minipage{0.45\textwidth}
  \includegraphics[width=\linewidth]{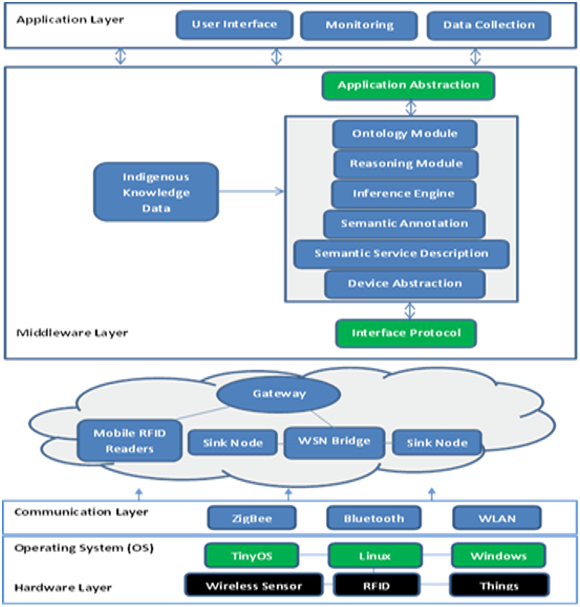}
  \caption{Overview of the middleware architecture}
\endminipage
\end{figure}

\section{Results and Discussion}
The study is expected to produce a semantic based data integration middleware that semantically represents and integrates heterogeneous data sources with indigenous knowledge based on a unified ontology for an accurate IoT-based drought forecasting system. With more integrated comprehensive services that are based on semantic interoperability, our approach makes a unique contribution towards improving the accuracy of drought prediction and forecasting systems.

%\begin{figure}
% use the \includegraphics command from the graphics package
% (N.B. put \usepackage{graphics} into your preamble then)
% for inclusion of your images. A sample call could read:
% \includegraphics{myimage}
%\vspace{2.5cm}
%\caption{This is the caption of the figure displaying a white eagle and
%a white horse on a snow field}
%\end{figure}

%

\end{document}